\documentclass[twocolumn,superscriptaddress,floatfix,prl]{revtex4-2}
\usepackage{graphicx}
\usepackage{color}
\usepackage{braket}
\usepackage{amsmath}

\newcommand{\PRB}{Phys. Rev. B }

\begin{document}
\title{Diagrammatic quantum Monte Carlo study of an acoustic lattice polaron}

\author{Thomas Hahn}
\affiliation{Faculty of Physics, Center for Computational Materials Science, University of Vienna,
A-1090 Vienna, Austria}
\author{Naoto Nagaosa}
\affiliation{RIKEN Center for Emergent Matter Science (CEMS),
2-1 Hirosawa, Wako, Saitama, 351-0198, Japan}
\affiliation{Department of Applied Physics, The University of Tokyo 7-3-1 Hongo, Bunkyo-ku,
Tokyo 113-8656, Japan}
\author{Cesare Franchini}
\affiliation{Faculty of Physics, Center for Computational Materials Science, University of Vienna,
A-1090 Vienna, Austria}
\affiliation{Dipartimento di Fisica e Astronomia, Universit\`a  di Bologna, 40127 Bologna, Italy}
\author{Andrey S. Mishchenko}
\affiliation{RIKEN Center for Emergent Matter Science (CEMS),
2-1 Hirosawa, Wako, Saitama, 351-0198, Japan}

\begin{abstract}
We present the first approximation free diagrammatic Monte Carlo study of a lattice polaron
interacting with an acoustic phonon branch through the deformation potential.
Weak and strong coupling regimes are separated by a self-trapping region where quantum resonance between
various possible lattice deformations is seen in the ground state properties, spectral function,
and optical conductivity.
The unique feature of such polaron is the interplay between long- and short wavelength acoustic vibrations
creating a composite phonon cloud and leading to persistent self-trapping due to the existence of
multiple quasi-stable states.
This results in a spectral response whose structure is much more complex than in any of the
previously considered polaron models.
\end{abstract}

\maketitle

The fundamental conception of a polaron, an electron interacting with phonons and increasing its mass
by dragging the accompanying phonon cloud with it, was formed long ago starting from the works of
Landau \cite{Landau_33} and Pekar \cite{Pekar_46}.
Depending on the type of phonon involved as well as the underlying mechanism used to model the
electron-phonon coupling, it is possible to distinguish between optical \cite{Frohlich_54} vs.
acoustic \cite{Bardeed_Shockley_50,SumiToyozawa_73} and lattice \cite{Holstein_59} vs.
continuum \cite{LLP_53} polarons.
This has led to an enormous variety of different interaction types
\cite{Barisic_S_70,Barisic_S_72,Pekar_79,SSH_79,SSH_88,SSH,Goulko_16a}.
Especially, optical polaron models have been studied extensively over the years.
The nearly constant spectrum $\omega({\bf q}) \approx \omega_0 > 0$ of optical phonons makes them
accessible to approximate approaches \cite{Ber_06,GooBer_08,GooMisBer_11}, approximation free
numerical techniques
\cite{Wellein_97,Kornil_98,Kornil_99,BoTruBa_99,Fehske_00,KuTruBo_02,Spenser_05,Hague_06,ViBoTru_10},
and even analytic methods \cite{Gerlach_Lowen_91}.
Acoustic polaron models, on the other hand, have received much less attention.
Due to their gapless and dispersive phonon spectrum, many techniques, which work well for the optical
case, either fail \cite{Gerlach_Lowen_91} or are effective only at weak coupling \cite{Marsiglio_11}.

In general, a typical polaron model describes an electron interacting with a single phonon branch ($\hbar=1$):
\begin{equation}
\begin{split}
\hat{H} = &\sum_{{\bf k}} \epsilon_{\bf k} c_{{\bf k}}^{\dagger} c_{{\bf k}} +
\sum_{\bf q} \omega({\bf q}) b_{{\bf q}}^{\dagger} b_{{\bf q}} + \\
&\frac{1}{\sqrt{N}}
\sum\limits_{{\bf k},{\bf q}} V({\bf q},{\bf k}) \; c_{{\bf k}+{\bf q}}^{\dagger} c_{{\bf k}}
(b^{\dagger}_{-{\bf q}}+ b_{{\bf q}}) \; .
\end{split}
\label{Ham}
\end{equation}
Here, the operator $c_{{\bf k}}^{\dagger}$/$b_{{\bf q}}^{\dagger}$ creates an electron/phonon with
momentum $\mathbf{k}$/$\mathbf{q}$.
The Hamiltonian is fully determined by the electron dispersion $\epsilon_{\bf k}$, the phonon frequency
$\omega({\bf q})$, and the interaction vertex $V({\bf q},{\bf k})$.
In lattice models, $N$ specifies the number of unit cells and sums over momenta are restricted to the
first Brillouin zone (BZ).

Every polaron eigenstate $\ket{\nu, \mathbf{k}}$ of Eq.~(\ref{Ham}) with energy
$E_\nu(\mathbf{k})$ can
be written as a linear combination of states in which an electron with momentum
${\bf k}-{\bf q}_1-\ldots-{\bf q}_n$ is accompanied by $n$ phonons with momenta
${\bf q}_1,\ldots,{\bf q}_n$:
\begin{equation}
\ket{\nu, \mathbf{k}}  =
\sum_{n=0}^{\infty} \sum_{{\bf q}_1 \ldots {\bf q}_n} \!\!
\Theta^{\nu,{\bf k}}_{{\bf q}_1 \ldots {\bf q}_n} c_{{\bf k}-{\bf q}_1-\ldots-{\bf q}_n}^{\dagger}
b^{\dagger}_{{\bf q}_1} \ldots b^{\dagger}_{{\bf q}_n} \ket{\mbox{\O}} \; .
\label{exp_ps}
\end{equation}
$\ket{\mbox{\O}}$ denotes the electron and phonon vacuum and $\Theta^{\nu,{\bf k}}_{{\bf q}_1 \ldots {\bf q}_n}$ are expansion coefficients.
For a fixed polaron momentum $\mathbf{k}$, we can identify the eigenstate $\ket{\nu=0,{\bf k}}$ with the lowest energy
$E_{0}({\bf k})$.
The ground state (GS) $\ket{0, \mathbf{k}_{\mathrm{GS}}}$, which is usually
located at $\mathbf{k}=\mathbf{k}_{\mathrm{GS}}=0$, has an energy
$E_{\mathrm{GS}}=E_{0}({\bf k}={\bf k}_{\mathrm{GS}})$ and is
further characterized by an effective mass
$m^* = (d^2 E_{0}(\mathbf{k}_{\mathrm{GS}}) / d {\bf k}^2)^{-1}$  and the
structure of its phonon cloud.
To describe the latter, one can use the probabilities of finding exactly $n$-phonons in the GS
$Z(n)=\sum_{{\bf q}_1 \ldots {\bf q}_n}
\left| \Theta^{0,{\bf k}=\mathbf{k}_{\mathrm{GS}}}_{{\bf q}_1 \ldots {\bf q}_n} \right|^2$
and their average number $\langle N_{\mathrm{ph}}\rangle  = \sum_n n Z(n)$.
More information on the excited spectrum is contained in the spectral function
$A_{\bf k}(\omega) = \sum_{\nu} \delta(\omega - E_{\nu}({\bf k}))
\left| \bra{\mbox{\O}} c_{\bf k} \ket{\nu, \mathbf{k}} \right|^2$,
which has poles (sharp peaks) at energies of stable (metastable) states of the polaron.
Furthermore, the weight of the zero-phonon line of the GS $Z_0=Z(0)$ can be retrieved from
$A_{{\bf k}=\mathbf{k}_{\mathrm{GS}}}(\omega) = Z_0 \delta(\omega -E_{\mathrm{GS}}) +
A_{\mathrm{inc}}(\omega)$ as the weight of the sharp $\delta$-functional peak
at the bottom of the spectrum while the smooth incoherent part $A_{\mathrm{inc}}(\omega)$
is located at $\omega>E_{\mathrm{GS}}$.

Gross features of optical polaron models with phonon dispersion $\omega({\bf q})=\omega_0>0$
and interaction vertex $V({\bf q},{\bf k})=V({\bf q})$ are well understood.
The electron-phonon coupling strength in these models can often be quantified with a single parameter
\begin{equation}
\widetilde{\lambda} = \sum_{\bf q} \; \frac{2 \, |V({\bf q})|^2}{W \, \omega({\bf q})} \; ,
\label{lam}
\end{equation}
where $W$ is the width of the electronic band.
Polaron states for $\widetilde{\lambda} \ll 1$ and $\widetilde{\lambda} \gg 1$ show profound
differences.
In the weak coupling regime (WCR) at $\widetilde{\lambda} \ll 1$, the GS is light
$m^*/m^*(\widetilde{\lambda}=0) \approx 1$
with a slightly distorted lattice $\langle N_{ph}\rangle \ll 1$ around the electron, whereas in the strong
coupling regime (SCR) at $\widetilde{\lambda} \gg 1$, the GS is heavy
$m^*/m^*(\widetilde{\lambda}=0) \gg 1$ because of a large phonon cloud
$\langle N_{ph}\rangle \gg 1$ surrounding the electron.

At $\widetilde{\lambda} = \widetilde{\lambda}_{\mathrm{cr}} \approx 1$, one may observe
the self-trapping (ST) phenomenon.
ST, as it was defined in the pioneer works \cite{Landau_33,Rashba_82,Ueta_86,Iosel_92},
has nothing to do with real trapping of a polaron in direct space.
Instead, it refers to an apparent quantum resonance between WCR and SCR.
The first indicator for ST is found in the shape of the phonon cloud.
For the lattice polaron with optical phonons it is a mixture of WCR and SCR states which manifests itself
as two peaks in the distribution $Z(n)$ at $n=0$ and $n>1$.
The second indicator is the clear avoided crossing behavior of the GS and first excited state (FES),
which can be stable in a model with optical phonons, in the spectral response.
As $\widetilde{\lambda}$ approaches the critical value $\widetilde{\lambda}_{\mathrm{cr}}$
from below, the GS and FES peak get closer and closer together, when finally they exchange
their properties at $\widetilde{\lambda} = \widetilde{\lambda}_{\mathrm{cr}}$, marking the
WCR to SCR crossover.
These two fingerprints are very spectacular when the optical phonon frequency $\omega_0$ is sufficiently
small $\omega_0 \ll W$, but they are hardly observed at large $\omega_0$
\cite{RP_02,Hohen_04}.

The phonon spectrum of the acoustic polaron model dramatically changes the situation
because the energy gap $\omega_0 > 0$, inherent to optical phonons, is missing and there is no
\textit{a priori} knowledge whether the Debye
frequency $\Omega_0$ of the acoustic phonon dispersion can be a good substitute.
It so happened that all previous non-perturbative studies were restricted to a simplified continuum
model \cite{Toyozawa_61,SumiToyozawa_73,Peeters_Devreese_85,Wang_98,Fantoni_12,Van_Houcke_15}
which, as we show below, has led to strange and counterintuitive conclusions.
The electronic band in the continuum model is approximated by $\epsilon_{\bf k}=k^2/2$ while the
phonon dispersion and interaction vertex are set to $\omega({\bf q}) = v_s q$ and
$V({\bf q}) \sim \sqrt{q}$, $v_s$ is a sound velocity.
A momentum cutoff $q \le k_0$ is introduced to establish the maximal Debye frequency of the phonon
spectrum $\Omega_0=v_s k_0$.
Surprisingly, it was shown \cite{SumiToyozawa_73,Toyozawa_61,Peeters_Devreese_85,Van_Houcke_15}
that the WCR to SCR crossover is sharper for {\it larger} values of $\Omega_0$.
This is clearly opposite to what is found in optical polaron models, where the crossover is sharper for
{\it smaller} $\omega_0$.
Studies based on approximate methods
\cite{SumiToyozawa_73,Toyozawa_61,Peeters_Devreese_85,Wang_98,Fantoni_12} even predicted
the possibility of an abrupt phase transition.
But exact diagrammatic Monte Carlo (DMC) calculations \cite{Van_Houcke_15} were able to
show that the crossover is indeed smooth
and thus supporting the opinion \cite{Peeters_82} that the abrupt transition is an artefact of the used
approximations rather than an intrinsic property of the model.

\begin{figure}[b]
\begin{center}
\includegraphics[width=8.5cm]{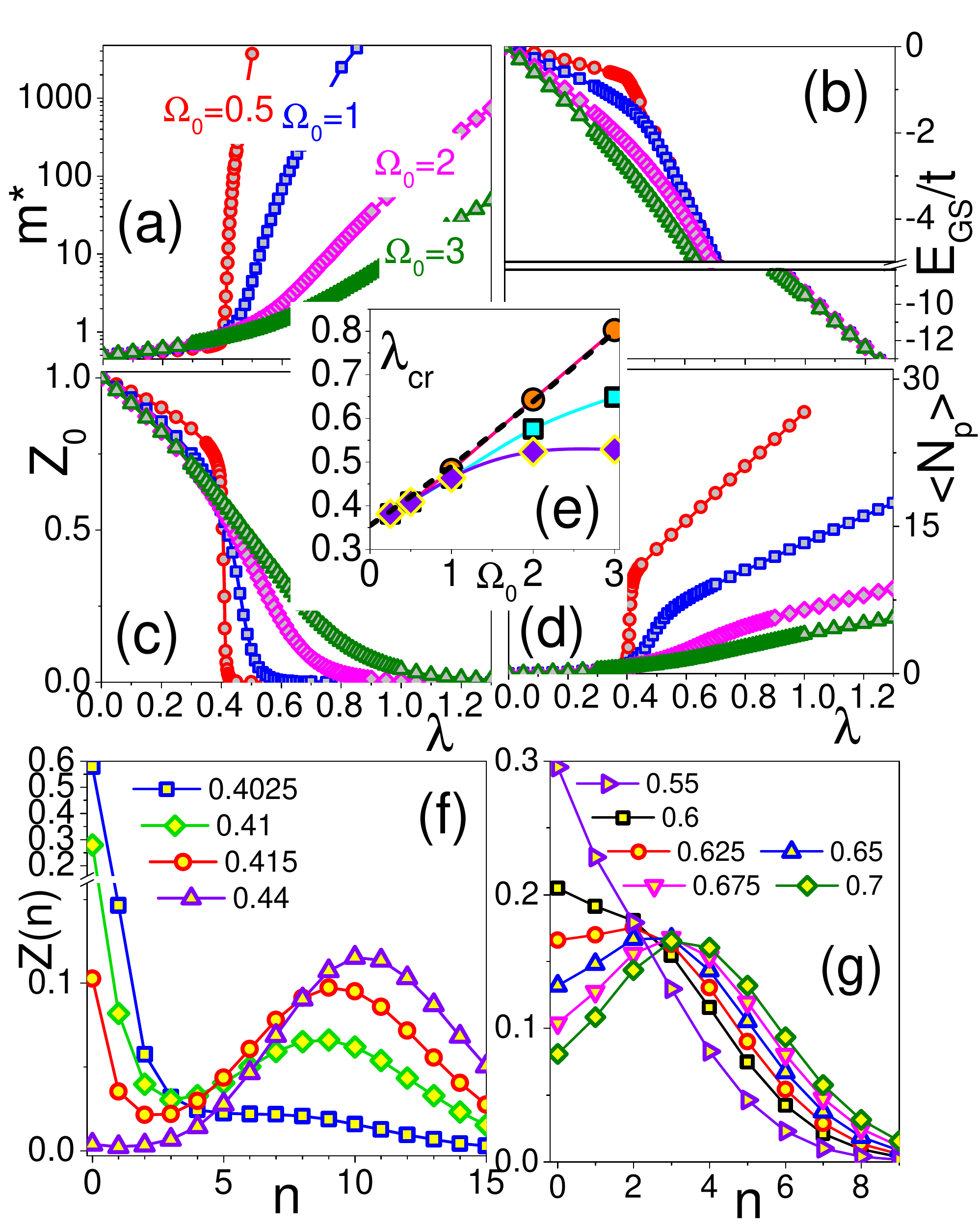}
\end{center}
\caption{\label{fig:fig1}
(a-d) Polaron ground state properties for $\Omega_0=0.5$ (red circles),
$\Omega_0=1$ (blue squares), $\Omega_0=2$ (magenta rhomboids), and $\Omega_0=3$ (olive triangles):
(a) effective mass $m*$, (b) ground state energy $E_{\mathrm{GS}}$, (c) quasiparticle weight $Z_0$, and
(d) average phonon number $\langle N_{\mathrm{ph}}\rangle$ in the phonon cloud.
(e) $\Omega_0$-dependence of the critical values $\lambda_{\mathrm{cr}}$
determined from the condition $d^2Z_0/d\lambda^2=0$ (rhomboids),
Min[$d^2E_{\mathrm{GS}}/d\lambda^2$] (squares), and
$d^2 \langle N_{\mathrm{ph}}\rangle /d\lambda^2=0$ (circles).
Dashed line shows extrapolation to zero Debye frequency $\Omega_0 \to 0$.
(f-g) Distribution $Z(n)$ of the number $n$ of phonons in the GS for different $\lambda$ (in legend)
for (f) $\Omega_0=0.5$ and (g) $\Omega_0=2$.
}
\end{figure}

In our opinion, a detailed study of an acoustic lattice polaron could help to shed some more light on the
nature of acoustic polarons and their connection to optical models.
Especially, approximation free calculations like DMC are able to provide reliable information for
the most interesting parameter region, i.e. the WCR to SCR crossover.
There is also growing evidence that realistic calculations require a thorough treatment of
acoustic polarons in the lattice \cite{Giustino_17,Giustino19}.
Experiments on coherent phonon generation in systems close to phase transitions, e.g charge-ordering
\cite{LCMO_05} or metal-insulator \cite{NNO_09} transitions, have measured significant coupling between
electrons and acoustic phonons.
In addition, scattering of charge carriers on acoustic phonons strongly influence their mobility in
many compounds, e.g in graphene \cite{Perebeinos_10}, solar cell materials \cite{Mali_17}, and organic
substances \cite{Acetylene_87,AcOrg_09}.
Very recently, experiments and first principle calculations have suggested that electron-acoustic phonon
coupling via the deformation potential is responsible for the formation of highly localized
(self-trapped) polarons in Cs\textsubscript{2}AgBiBr\textsubscript{6} \cite{Wu_21}.

In the present Letter, we study the ground and excited states of an acoustic lattice polaron model
using approximation free DMC \cite{PS98,MPSS,OCFr,Hahh} and stochastic optimization consistent
constraints \cite{MPSS,Julich,Goulko_16b} analytic continuation methods.
It will be shown that the properties of the acoustic lattice polaron are different from previously
studied model systems and it therefore constitutes a separate object in the vast family of
various polaron phenomena.

We consider a three dimensional (3D) primitive
cubic lattice with the corresponding standard tight binding dispersion for the electron
\begin{equation}
\epsilon({\bf k}) = 2t \sum_{i}^{x,y,z} \left[ 1-\cos(k_i) \right] \; , \; -\pi \le k_i < \pi \;,
\label{espe}
\end{equation}
($t$ is the nearest-neighbor hopping amplitude) and an acoustic phonon spectrum due to
nearest-neighbor force constants and with Debye frequency $\Omega_0$
\begin{equation}
\omega({\bf q}) = \Omega_0 \sqrt{ \sum_{i}^{x,y,z} \sin^2 \left\{ q_i/2 \right\} } \; , \; -\pi \le q_i < \pi \;.
\label{phon}
\end{equation}
Estimating the deformation potential
$D = \sigma {\rm div} ({\bf u}({\bf r}))  = \sigma \sum_{i}^{x,y,z} \partial u_i/\partial x_i$,
as the energy proportional to the change of the unit cell volume
$D \sim \sum_i^{x,y,z} \mid  {\bf u}_i({\bf r}+\hat{\bf a}_i)  - {\bf u}_i({\bf r}) \mid$,
where $\hat{\bf a}_i$ is a primitive unit vector of the 3D cubic lattice and
${\bf u}_i({\bf r}) \sim {\hat{\bf a}}_i\sum_{\bf q} e^{i{\bf qr}} \left( b_{q} + b_{-q}^{\dagger} \right) / \sqrt{\omega({\bf q})}$
is a displacement due to the acoustic phonon mode, one arrives at the interaction vertex
\begin{equation}
V({\bf q}) =
\gamma
 \sum_{i}^{x,y,z} \sin \left\{ \left| q_i/2 \right| \right\}
 \left[ \sum_{i}^{x,y,z} \sin^2 \left\{ q_i/2 \right\} \right]^{-1/4} \; .
\label{dpoten}
\end{equation}
As is usually the case, we define a dimensionless coupling constant
$\lambda = \gamma^2 / (6 t \Omega_0)$, such that $\widetilde{\lambda} \approx 2.5 \lambda$ [see
Eq.~(\ref{lam})].
The lattice constant is set to $a=1$ and $t=1$ throughout.

Figures~\ref{fig:fig1}(a)--\ref{fig:fig1}(d) show the ground state properties (effective mass, energy,
quasiparticle weight and average number of phonons) as functions of the coupling strength $\lambda$.
In contrast to the acoustic continuum polaron, the transition from WCR to SCR is sharper for smaller
values of the Debye energy cutoff $\Omega_0$.
This behavior is similar to what is observed in optical lattice models, where the crossover is more abrupt
for smaller ratios $\omega_0/W$.
Moreover, it is interesting to note that the value of
$\widetilde{\lambda}_{\mathrm{cr}} \approx 2.5 \lambda_{\mathrm{cr}}$
in the adiabatic limit is in good agreement in both optical and acoustic lattice models.
In Fig.~\ref{fig:fig1}(e), we show $\lambda_{\mathrm{cr}}(\Omega_0)$ determined from derivatives of
various GS properties with respect to $\lambda$ (see the caption for more details).
Although the values of $\lambda_{\mathrm{cr}}(\Omega_0)$ slightly depend on the kind of derivative, this
discrepancy becomes negligible at small values of $\Omega_0$ allowing us to extrapolate to the adiabatic
limit $\lambda_{\mathrm{cr}}(\Omega_0 \to 0) = 0.355 \pm 0.01$.
Using Eq.~(\ref{lam}), we can compare this value for the acoustic polaron
$\widetilde{\lambda}_{\mathrm{cr}}^{\mathrm{ac}}(\Omega_0 \to 0) = 0.895 \pm 0.03$ to the one obtained for
the optical polaron $\widetilde{\lambda}_{\mathrm{cr}}^{\mathrm{op}}(\Omega_0 \to 0) = 0.9$
\cite{ShinozukaTo_79,MisImpu_09}.

\begin{figure}[t]
\begin{center}
\includegraphics[width=8.5cm]{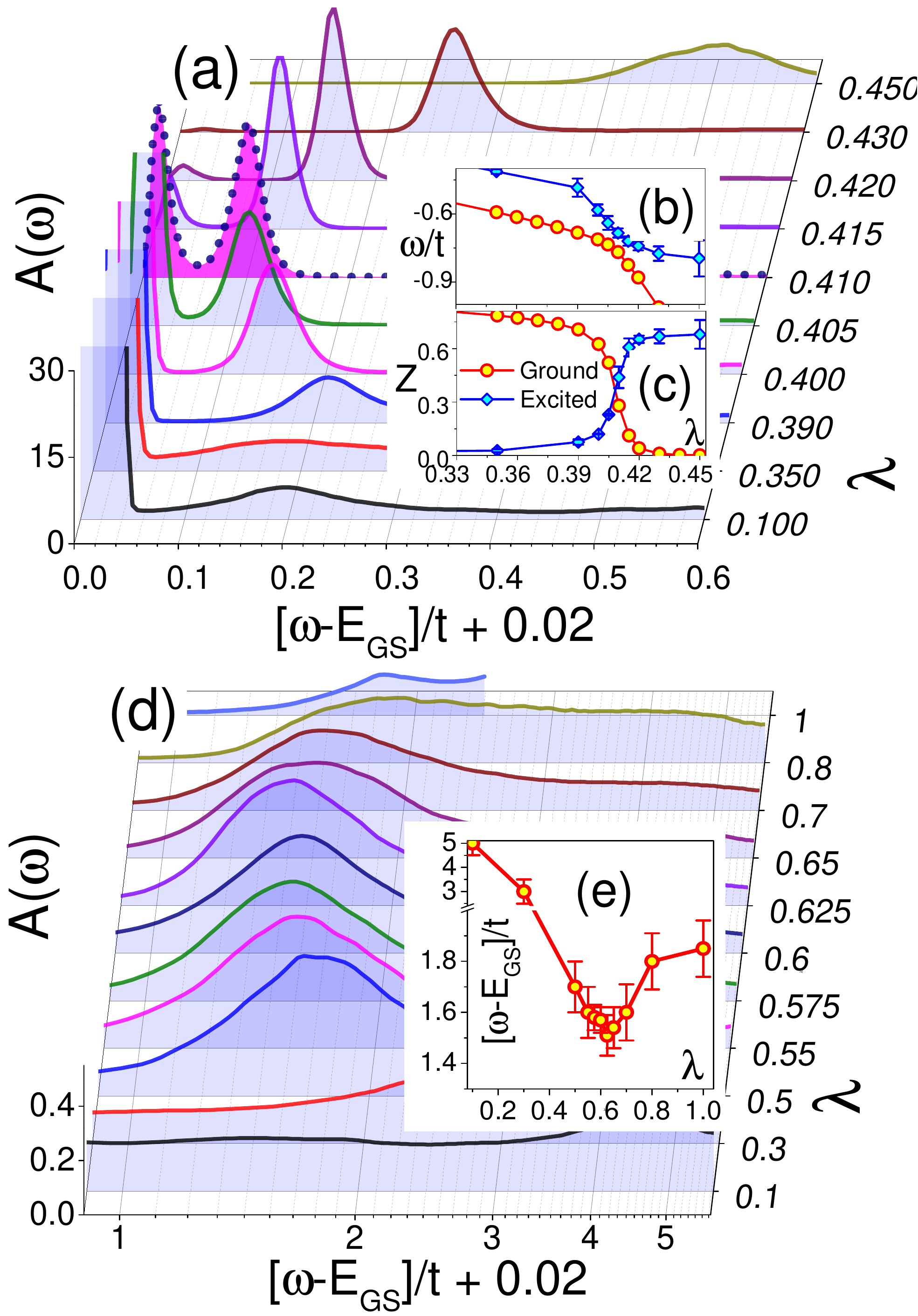}
\end{center}
\caption{\label{fig:fig2}
(a-c) $\Omega_0=0.5$: (a) Spectral functions $A(\omega)/\kappa$
[$\kappa$ with increasing $\lambda$ is 0.006, 0.02, 0.08, 0.12, 0.23, 0.44, 0.61, 0.65, 0.67, 0.68],
(b) energy, and (c) weight of the GS (red circles) and FES (blue squares) peaks.
The spectrum at $\lambda_{\mbox{\scriptsize cr}} \approx \lambda=0.41$ is highlighted by a dotted line.
(d-e) $\Omega_0=2$: (d) Incoherent part of spectral functions $A(\omega)/\kappa$
[$\kappa$ with increasing $\lambda$ is 0.04, 0.21, 0.45, 0.6, 0.75, 0.8, 0.8, 0.9, 0.95, 0.65, 0.2] and
(e) energy of the excited state peaks counted from $E_{\mathrm{GS}}$.
}
\end{figure}

The general features of the phonon distribution function $Z(n)$ in Figs.~\ref{fig:fig1}(f)
and~\ref{fig:fig1}(g) are very similar to the optical polaron case \cite{GooMisBer_11,QDST_01,RP_02}.
In the WCR, the free electron contribution to the expansion of the polaron GS in Eq.~(\ref{exp_ps})
is the dominant one and so $Z(n=0) \approx 1$ while $Z(n>0) \approx 0$.
In the SCR, $Z(n)$ resembles a Poisson distribution with its mean located at
$\approx 13.47 \, \lambda/\Omega_0$.
For the nearly adiabatic case $\Omega_0=0.5$ [Fig.~\ref{fig:fig1}(f)], $Z(n)$ develops a two-peak
structure close to the critical coupling $\lambda_{\mathrm{cr}}\approx 0.41$.
This self-trapping indicator is not seen in the nearly anti-adiabatic case $\Omega_0=2$
[Fig.~\ref{fig:fig1}(g)], where we always observe a single peak.

It is thus no surprise that for $\Omega_0=0.5$, in agreement with the resonance behavior of $Z(n)$,
the spectral function in Fig.~\ref{fig:fig2}(a) demonstrates the avoided crossing phenomenon representative
for self-trapping.
At $\lambda=0.1$, the FES is still well separated $\approx 0.2t$ from the GS, which contains nearly all of
the weight, i.e. $Z_0 \approx 1$.
With increasing coupling, this gap becomes smaller and the weight is slowly transferred to the FES, as can be seen from Figs.~\ref{fig:fig2}(b) and~\ref{fig:fig2}(c).
At the crossover point $\lambda = \lambda_{\mathrm{cr}}$, the energy gap between GS and FES is minimal and
their weights become equal.

\begin{figure}[]
\begin{center}
\includegraphics[width=8.7cm]{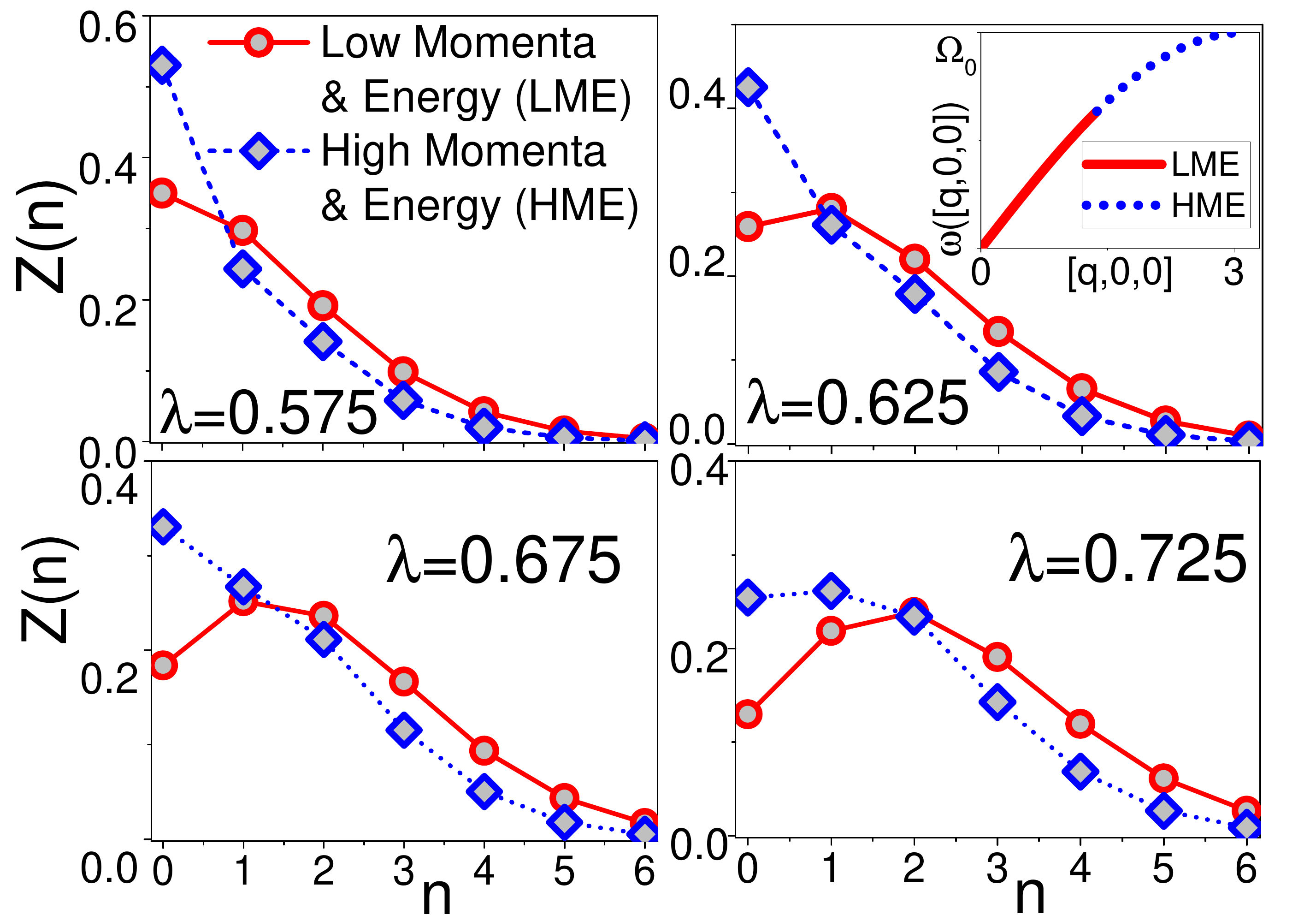}
\end{center}
\caption{\label{fig:fig3}
Distribution of the number $n$ of phonons in the ground state for $\Omega_0=2$. Phonons are grouped into
LME/HME (circles/rhomboids).
Inset in the upper right corner shows the division into LME and HME.
}
\end{figure}

On the other hand, it seems surprising that we can see the avoided crossing hybridization behavior for
$\Omega_0=2$ [Figs.~\ref{fig:fig2}(d) and~\ref{fig:fig2}(e)], since there is no corresponding two-peak
structure in the phonon distribution function $Z(n)$ [see Fig.~\ref{fig:fig1}(g)].

\begin{figure}[]
\begin{center}
\includegraphics[width=8.5cm]{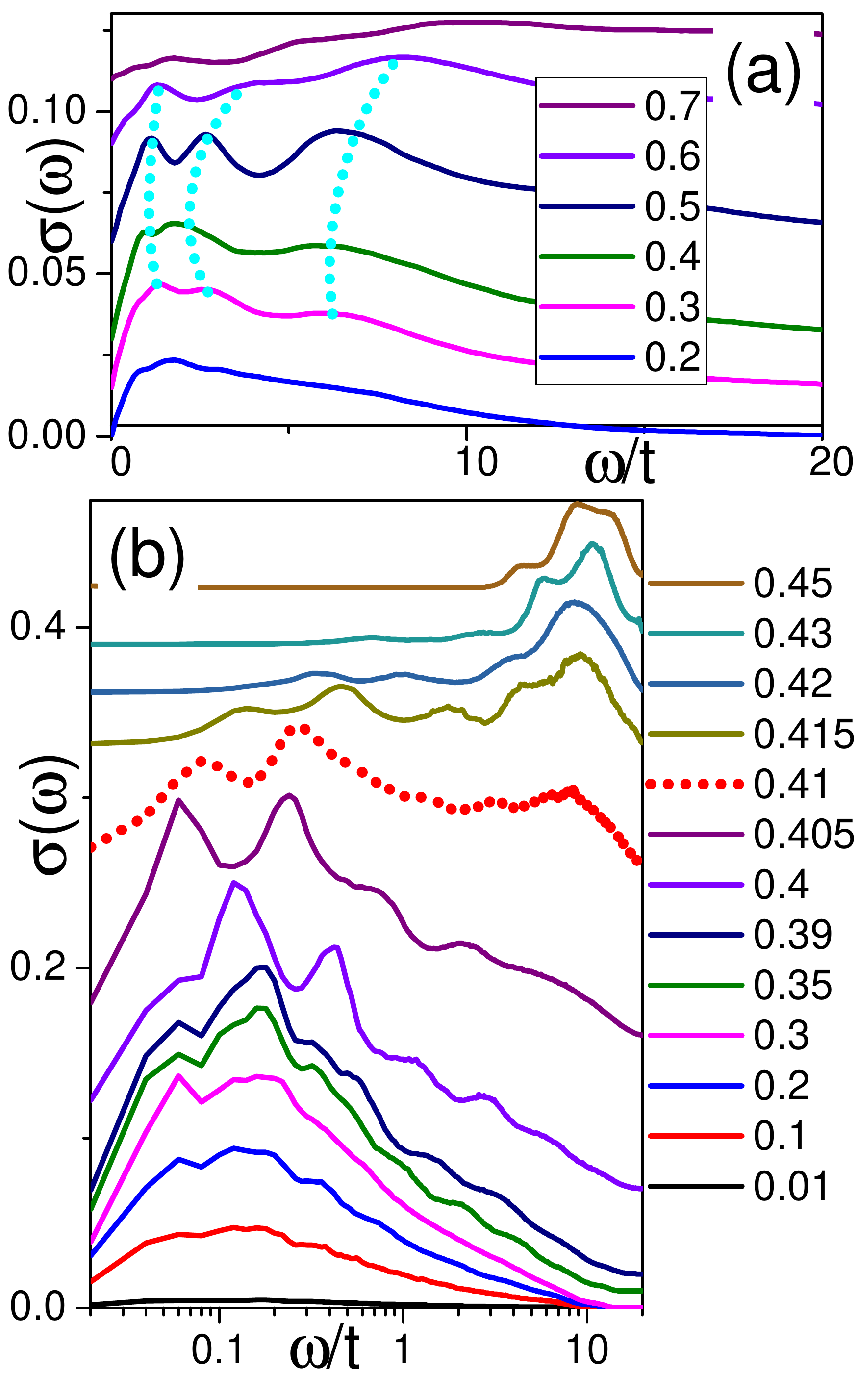}
\end{center}
\caption{\label{fig:fig4}
Optical conductivity $\sigma(\omega)+\chi$ for different $\lambda$ given in the legend.
$\chi$ is an artificially added onset to discern the curves for different values of $\lambda$:
(a) $\Omega_0=2$
[$\chi$ with increasing $\lambda$ is 0, 0.015, 0.03, 0.06, 0.09, 0.11] and
(b) $\Omega_0=0.5$
[$\chi$ with increasing $\lambda$ is 0, 0, 0, 0, 0.01, 0.02, 0.07, 0.16, 0.26, 0.33, 0.36, 0.39, 0.42].
Dotted blue lines in (a) follow peaks to guide the eye.
The spectrum in (b) at $\lambda_{\mathrm{cr}} \approx \lambda=0.41$ is highlighted by the
dotted red line.
}
\end{figure}

However, the unique feature of the acoustic lattice polaron is the presence of two distinctive
groups of phonons whose properties are significantly different.
Considering the dispersion in Eq.~(\ref{phon}) and the interaction vertex in Eq.~(\ref{dpoten}), one can
distinguish between (i) low momenta and energy (LME) phonons close to the $\Gamma$ point and (ii) high momenta and energy (HME) phonons near the BZ boundary.
We tentatively divide the phonons into LME/HME by using a sphere with radius $q_r =(3\pi^2)^{1/3}$
and half the volume of the BZ [see inset in Fig.~\ref{fig:fig3}].
Phonons with momenta inside the sphere belong to LME otherwise to HME.
Figure~\ref{fig:fig3} shows their different behavior for $\Omega_0=2$.
The LME phonon cloud transfers from WCR to SCR at $\lambda \approx 0.6$, while the HME cloud makes
the crossover at a considerably stronger coupling  $\lambda \approx 0.725$.
Depending on $\lambda$, it is thus possible to observe three qualitatively different situations in the GS:
(i) LME and HME are both in the WCR, (ii) LME is in the SCR while HME is still in the WCR and
(iii) LME and HME are both in the SCR.
The forth combination, (iv) LME is in the WCR and HME is in the SCR, is not seen in the
GS but it might be realized in some excited states.
Furthermore, the very nature of acoustic phonons implies that there is always a group of soft phonons
present, namely LME, which are responsible for the manifestations of the self-trapping phenomenon even for
large Debye frequencies $\Omega_0$, see Figs.~\ref{fig:fig2}(d) and~\ref{fig:fig2}(e).
For small Debye frequencies $\Omega_0=0.5$, both LME and HME show a two-peak structure while the
number of phonons $n$ at maximal $Z(n)$ is, indeed, larger for LME (See Fig.~S1 in Supplementary
material \cite{Supplement}).

The anomalously rich structure of the phonon cloud leads to the existence of multiple competing states.
For example, a third competing state is clearly seen in the spectral function $A(\omega)$ for
$\Omega_0=0.5$ (see Fig.~S2 in Supplementary material \cite{Supplement}).
To reveal more details on the excited spectrum, one can look at the zero temperature optical conductivity
(OC) $\sigma(\omega)$, since the symmetry selection rules for $A(\omega)$ and $\sigma(\omega)$ are different
\cite{GooMisBer_11,OBarisic_04}.
We are aware of only one pioneering variational study which investigated the optical response of the
acoustic continuum polaron \cite{Peeters_Impedance_87}.
Figure~\ref{fig:fig4} presents approximation free results for $\sigma(\omega)$ for our lattice model.
Numerous excited states can be identified not only for $\Omega_0=0.5$ (Fig.~\ref{fig:fig4}(b)) but
even for the $\Omega_0=2$ case (Fig.~\ref{fig:fig4}(a)).
This is possibly a reflection of the unique composite phonon cloud.
In this regard, other numerical or analytical techniques, like the double phonon cloud approach
\cite{DeFilippis_12} or the momentum average approximation \cite{GooMisBer_11}, could help with
a better understanding of the shape of the OC curves and how they relate to other properties of the
polaron.

In conclusion, we tried to show that the lattice polaron interacting with an acoustic phonon branch via the
deformation potential is an object whose complexity is going far beyond the previously studied optical
and acoustic continuum polarons.
Its most interesting feature is the existence of two groups of phonons which can interact with the
quasiparticle in different ways.
These two phonon clouds are responsible for the rich structure in the optical
response showing multiple excited states and for the persistent self-trapping phenomenon even at
large values of the Debye frequency.
Furthermore, the sharpness of the crossover between weak and strong coupling  has a different
dependence on the Debye/cut-off frequency in the acoustic continuum polaron compared
to our lattice model.
This shows that these two models are not simply two limiting cases of each other but that they describe
very different objects.

{\it Acknowledgements}.
We are grateful to O. Bari\v{s}i\'{c} and S. N. Klimin for fruitful discussions.
This work was supported by JST CREST Grant Number JPMJCR1874, Japan and by the joint FWO-FWF Grant
No. I 2460-N36 and Grant No. I 4506. The computational results
presented have been achieved in part using the Vienna Scientific Cluster (VSC).

\end{document}


\title{Supplemental material for ''Diagrammatic quantum Monte Carlo study of an acoustic lattice polaron''}

\author{Thomas Hahn}
\author{Naoto Nagaosa}
\author{Cesare Franchini}
\author{Andrey S. Mishchenko}

\maketitle

Figure S1 presents the distribution of the number $n$ of phonons in the ground state for $\Omega_0=0.5$.

\begin{figure}[htb]
\begin{center}
\includegraphics[width=8.5cm]{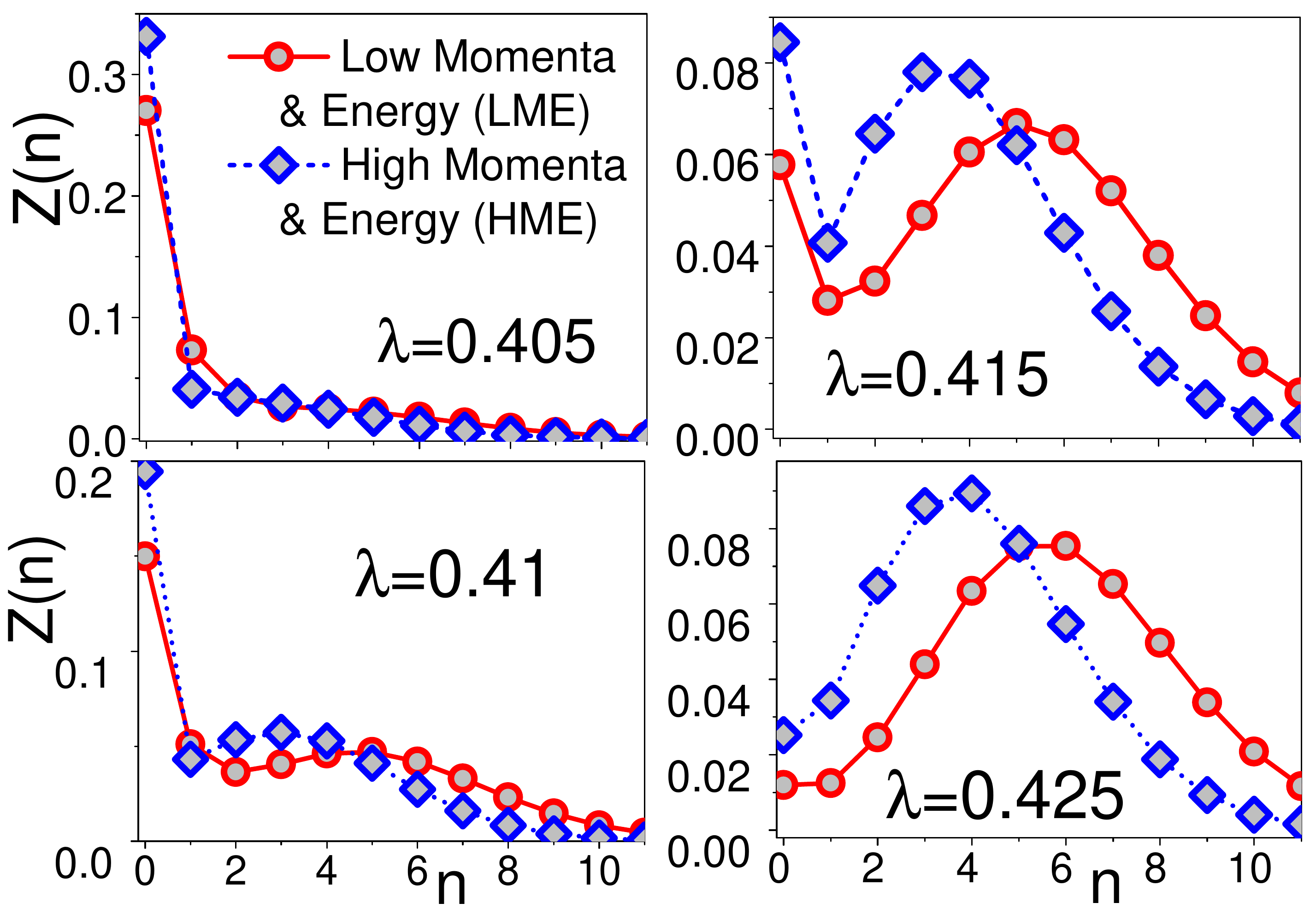}
\end{center}
\caption{\label{fig:fig1}
Distribution of the number $n$ of phonons in the ground state for $\Omega_0=0.5$.
Phonons are grouped into LME/HME (circles/rhomboids).
}
\end{figure}

Figure S2 presents the dependence of the spectral function $A(\omega)$ for $\Omega_0=0.5$
on the coupling constant $\lambda$
in the scale which allows to see the second excited state at $[\omega-E_{GS}]/t \approx 0.5$.

\begin{figure}[htb]
\begin{center}
\includegraphics[width=8.5cm]{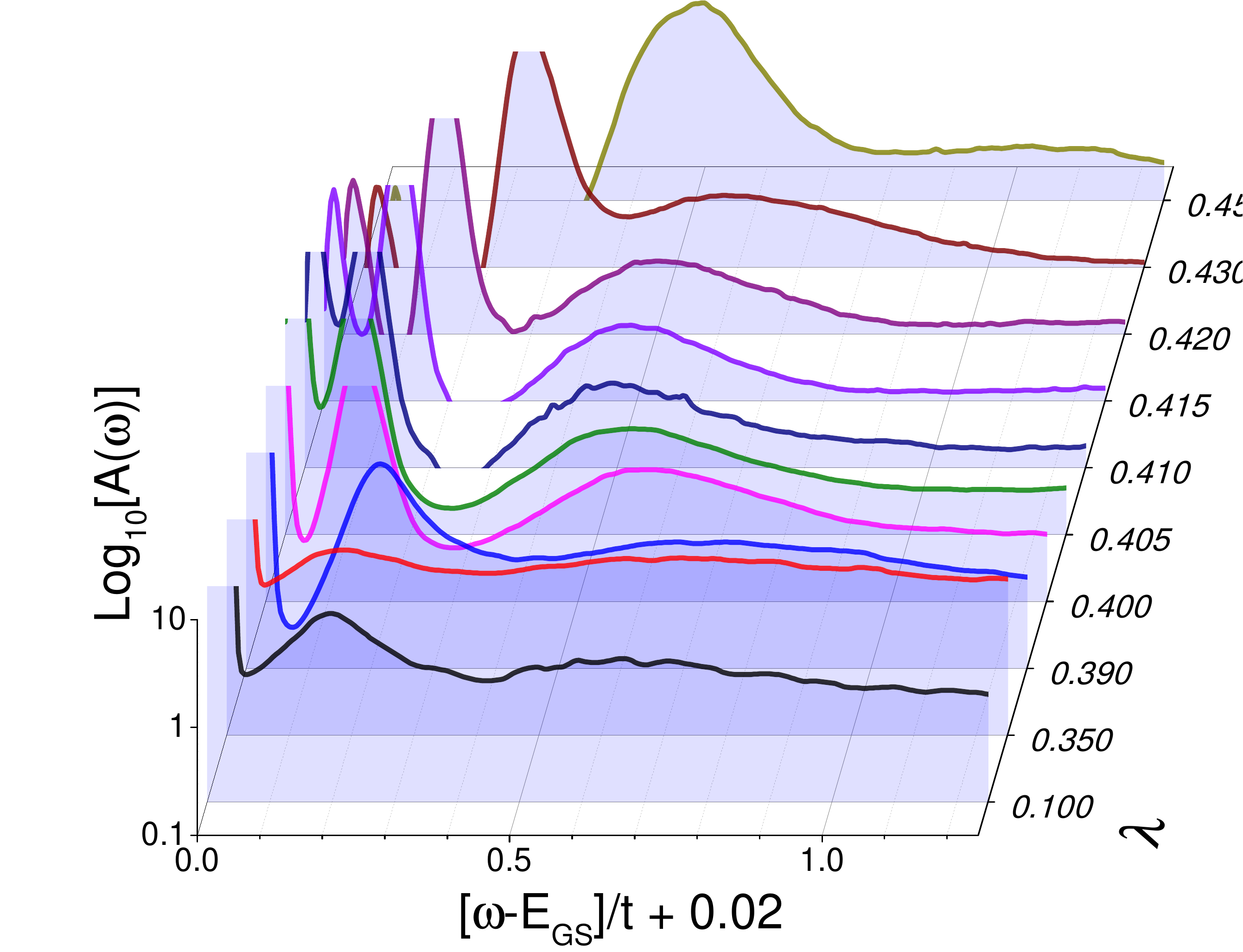}
\end{center}
\caption{\label{fig:fig2}
Spectral functions $A(\omega)/\kappa$ for $\Omega_0=0.5$ and different coupling strengths $\lambda$.
$\kappa$ with increasing $\lambda$ is
0.006, 0.02, 0.08, 0.12, 0.23, 0.44, 0.61, 0.65, 0.67, 0.68.
}
\end{figure}



